\documentclass{iopjournal}

\usepackage{amsmath}
\usepackage{amssymb}
\usepackage{array}
\usepackage{tabularx}
\usepackage{booktabs}
\usepackage{subcaption}
\usepackage{algorithm}
\usepackage{algpseudocode}
\usepackage{colortbl}
\usepackage{placeins}
\usepackage{svg}
\usepackage{microtype}

\begin{document}

\articletype{Paper} 

\title{Efficient Quantum Modular Reduction: Crandall reduction and its Fault-tolerant resource analysis}

\author{Changyeol Lee$^{1,\dagger}$\orcid{0009-0005-2052-5626}, Sungyeon Kook$^{1,\dagger}$\orcid{0009-0009-8689-6250}, Wooyeong Song$^1$\orcid{0000-0002-7127-5170}, Kwangil Bae$^1$\orcid{0009-0007-7001-671X}, Wonhyuk Lee$^1$ and IlKwon Sohn$^{1,*}$\orcid{0000-0001-5321-2427}}

\affil{$^1$Quantum Network Research Center, Korea Institute of Science and Technology Information, Daejeon 34141, Republic of Korea}

\affil{$^\dagger$These authors contributed equally to this work.}

\affil{$^*$Author to whom any correspondence should be addressed.}

\email{d2estiny@kisti.re.kr}


\abstract{Modular arithmetic is central to quantum algorithms for cryptographic problems, including Shor's algorithm and Grover-based cryptanalysis, with modular reduction contributing substantially to circuit cost. Pseudo-Mersenne moduli \(q=2^n-c\) allow classical Crandall reduction to replace division with folding and constant arithmetic, providing a structural opportunity for more efficient quantum modular reduction than Barrett reduction. We translate this advantage into a reversible quantum setting by deriving explicit folding and normalization conditions for \(2n\)-bit inputs. To the best of our knowledge, this constitutes the first exact reversible quantum circuit formulation of Crandall reduction. Based on this formulation, we develop two variants: Crandall reduction-1 is designed to minimize execution cost through one-step normalization, whereas Crandall reduction-2 uses two-step normalization to support a wider range of \(c\) with limited overhead. Logical resource estimates show that both variants require fewer qubits and lower T-count and T-depth than optimized folding Barrett reduction. At \(n=10\), Crandall reduction-1 reduces both T-count and T-depth by approximately \(46.9\%\) relative to optimized folding Barrett reduction. Surface-code analysis further shows that, at \(n=20\) under the Sparse Blossom decoder, the estimated runtimes of the two variants are \(30.05\,\mathrm{ms}\) and \(35.39\,\mathrm{ms}\), respectively, compared with \(53.77\,\mathrm{ms}\) for optimized folding Barrett reduction. These results demonstrate the practical value of exploiting modulus-specific arithmetic structure in fault-tolerant quantum circuit design.}
\section{Introduction}\label{sec:Introduction}
Designing efficient modular arithmetic circuits is an important challenge in quantum computing because arithmetic operations must be implemented reversibly~\cite{bennett1973logical}. Intermediate information cannot be discarded arbitrarily and must instead be retained or subsequently uncomputed, often increasing the required numbers of qubits and operations~\cite{paradis2024reqomp}. These constraints are particularly significant for modular operations, which generally require comparisons and conditional reduction or restoration to map outputs into a prescribed range~\cite{orts2024quantum,thapliyal2017quantum}. Consequently, optimizing modular reduction is important not only for improving circuit-level efficiency, but also for designing larger quantum algorithms that repeatedly invoke modular arithmetic and accurately estimating their implementation costs. This motivates the identification of modulus classes whose algebraic structure can simplify modular reduction and thereby reduce the cost of reversible implementations.

Because direct division is costly to implement reversibly, we consider classical modular reduction methods that replace it with operations such as multiplication by precomputed constants, bit shifts, and additions or subtractions. Montgomery reduction performs reduction in a transformed representation~\cite{montgomery1985modular}, whereas Barrett reduction approximates the quotient using a precomputed reciprocal~\cite{barrett1987implementing}. Among modulus classes that permit further specialization, pseudo-Mersenne moduli \(q=2^n-c\), where \(c\) is small, enable efficient modular arithmetic based on folding and multiplication by a small constant. The congruence \(2^n \equiv c \pmod q\) allows Crandall reduction to fold the high-order bits into the lower part using multiplication by \(c\) and addition~\cite{crandall1993method}. Compared with the quotient-approximation procedure used in Barrett reduction, this structure provides an opportunity to construct a lower-cost quantum modular reduction circuit. Realizing this potential in a reversible circuit, however, requires careful treatment of reversible folding, efficient quantum-classical multiplication by \(c\), and normalization of the folded result to \([0,q)\). We analyze the output range after two folding operations and derive the conditions under which one or two normalization steps are sufficient. Based on this analysis, we propose quantum Crandall reduction circuits specialized for pseudo-Mersenne moduli and demonstrate lower resource costs than the optimized folding Barrett reduction circuit proposed by Zhang et al.~\cite{zhang2025optimized}.

In addition, we analyze the efficiency of the proposed quantum modular reduction circuit in a fault-tolerant quantum computing environment. Based on the surface code, which has emerged as a leading quantum error-correcting framework due to its practical hardware constraints and robust scaling properties, we compare the resource requirements of existing modular reduction circuits and the proposed design. In the surface-code environment, the implementation cost of the non-Clifford T gate is dominant compared with Clifford gates, and T gates are supplied fault-tolerantly through a separate resource-intensive procedure such as a magic-state factory~\cite{fowler2012surface, fowler2019low}. Therefore, the reductions in \(T\)-count and \(T\)-depth go beyond reducing the number of gates by relaxing the logical error-rate requirement and potentially reducing the surface-code distance required to satisfy the target circuit success probability~\cite{litinski2019game, ha2024resource}. This leads to a reduction in logical-qubit and physical-qubit overhead, and at the same time mitigates the decoder processing time and decoder backlog for syndrome data, ultimately leading to a reduction in the total execution time of the quantum circuit.

In summary, this work makes two main contributions. First, we develop efficient quantum modular reduction circuits specialized for pseudo-Mersenne moduli \(q=2^n-c\). By adapting Crandall reduction to a reversible quantum setting, we present two circuit variants, Crandall reduction-1 and Crandall reduction-2, based on their normalization procedures. Second, we demonstrate that the proposed circuits achieve lower \(T\)-count and \(T\)-depth than existing Barrett reduction circuits, including optimized folding Barrett reduction. For example, at \(n=10\), Crandall reduction-1 reduces both the \(T\)-count and \(T\)-depth by approximately \(46.9\%\) compared with optimized folding Barrett reduction. Our fault-tolerant analysis further shows that the reduced \(T\)-depth can relax the logical error-rate requirement, lower the required surface-code distance, and shorten the overall execution time.
\section{Analysis of Crandall reduction for quantum circuit design}

This section presents the analytical basis for the proposed quantum Crandall reduction circuits. We first review classical Crandall reduction, then determine the required number of folding operations, and finally derive the conditions for one-step and two-step normalization. These results define the two circuit variants presented in Section 3.

\subsection{Classical Crandall reduction}

Crandall reduction is an efficient modular reduction technique for pseudo-Mersenne moduli \(q \in \mathbb{Z}_{\ge 2}\) of the form

\begin{equation}
q = 2^\ell-c,
\label{eq:1}
\end{equation}

where \(\ell\) and \(c\) are positive integers, with \(\ell\) denoting the bit-length parameter and \(c\) being small relative to \(2^\ell\)~\cite{crandall1993method}. The method relies on the congruence

\begin{equation}
2^\ell \equiv c \pmod{q}.
\label{eq:2}
\end{equation}

For an integer \(a\), we write
\begin{equation}
a = a_1 2^\ell + a_0.
\label{eq:3}
\end{equation}
Then,
\begin{equation}
a \equiv a_1 2^\ell + a_0
  \equiv c a_1 + a_0 \pmod{q}.
\label{eq:4}
\end{equation}

Thus, the upper part of \(a\) can be folded into the lower part by multiplying \(a_1\) by \(c\) and adding the result to \(a_0\). This folding step is repeated until the intermediate value becomes smaller than \(2q\). If the remaining value is still greater than or equal to \(q\), one final subtraction of \(q\) yields the reduced result. Algorithm~\ref{alg:crandall_reduction} summarizes the overall procedure.

\begin{algorithm}[h]
\caption{Crandall reduction}
\label{alg:crandall_reduction}
\begin{algorithmic}[1]
\Require \(a\), \(q = 2^\ell - c\)
\While{\(a \ge 2q\)}
    \State \(a_0 \gets a \mathbin{\&} (2^\ell - 1)\)
    \State \(a_1 \gets a \gg \ell\)
    \State \(a \gets c a_1 + a_0\)
\EndWhile
\If{\(a \ge q\)}
    \State \(a \gets a - q\)
\EndIf
\State \Return \(a\)
\end{algorithmic}
\end{algorithm}

For pseudo-Mersenne moduli of the form \(2^\ell-c\) with small \(c\), this folding structure avoids division and large-integer multiplication by using bit shifts, multiplication by the small constant \(c\), and addition/subtraction. When this special structure can be exploited, Crandall reduction can provide more efficient modular reduction than general-purpose methods such as Barrett reduction and Montgomery reduction~\cite{bosselaers1994comparison,bajard2005modular}.

For the quantum circuit considered in this paper, we use the classical Crandall reduction principle as a basis, but the circuit-level setting requires a fixed reduction structure. In particular, the modular reduction input is assumed to be a \(2n\)-qubit value produced by an \(n \times n\) multiplication, and the target modulus is taken as \(q = 2^n - c\). Unlike the classical procedure in Algorithm~1, whose folding loop terminates depending on the intermediate value, the number of folding operations in a quantum circuit is fixed in advance. Once the folding count is fixed, the range of the folded result also determines how many normalization steps are required to obtain an output in the canonical range. Therefore, in the following subsections, we analyze two design factors for the proposed circuit: the number of folding operations and the normalization conditions after folding.

\subsection{Folding-count analysis}

We compare different numbers of folding operations in terms of the intermediate-value range and the final normalization cost to determine the folding strategy used in the proposed circuit.

Let the input value be expressed as \(a=a_1 2^n+a_0\), where \(0\leq a_1,a_0<2^n\). One folding operation gives
\begin{equation}
a^{(1)}=ca_1+a_0<(c+1)2^n.
\end{equation}
For the second folding operation, the first folded value is decomposed as
\begin{equation}
a^{(1)}=a_1'2^n+a_0',
\end{equation}
where \(0\leq a_0'<2^n\) and \(0\leq a_1'\leq c\). The second folding operation then gives
\begin{equation}
a^{(2)}=ca_1'+a_0'\leq c^2+(2^n-1).
\end{equation}
The maximum numbers of conditional subtractions required for final normalization after one and two folding operations are, respectively,
\begin{equation}
N_1^{\max}
=
\left\lceil
\frac{(c+1)2^n}{q}
\right\rceil-1
=
\left\lceil
\frac{(c+1)2^n}{2^n-c}
\right\rceil-1
=
c+
\left\lceil
\frac{c(c+1)}{2^n-c}
\right\rceil.
\label{eq:5}
\end{equation}
and
\begin{equation}
N_2^{\max}
=
\left\lceil
\frac{2^n+c^2}{q}
\right\rceil-1
=
\left\lceil
\frac{2^n+c^2}{2^n-c}
\right\rceil-1
=
\left\lceil
\frac{c(c+1)}{2^n-c}
\right\rceil.
\label{eq:6}
\end{equation}

Under the condition \(c^2+2c\leq 2^n\), these expressions give \(N_1^{\max}=c+1\) and \(N_2^{\max}=1\), respectively. Thus, two folding operations reduce the maximum number of final corrections from \(c+1\) to a single conditional subtraction, while also restricting the intermediate value to at most \(n+1\) qubits.

A third folding operation further reduces the upper bound from \(2^n+c^2\) to \(2^n+c\). However, under the same condition, both two and three folding operations require at most one conditional subtraction for final normalization. Therefore, a third folding operation incurs one additional folding cost without further reducing the normalization cost. Among the cases considered in this work, two folding operations provide the most resource-efficient choice.

\subsection{Normalization conditions after two foldings}

In Section~2.2, the number of folding operations in the quantum Crandall reduction circuit was fixed to two. The remaining issue is to determine how many normalization steps are required for the two-fold result. Since \(a(2) \le c^2 + (2^n - 1)\), the required normalization count depends on the magnitude of the constant \(c\). In this subsection, we derive the admissible ranges of \(c\) for one-step and two-step normalization, which determine the corresponding circuit variants proposed in the following section.

We first analyze the case in which final normalization is performed using a single conditional subtraction. For the result $a(2)$ after two folding operations, one-step normalization is possible if $a(2) < 2q$. Since $a(2) \le c^2 + (2^n - 1)$ and $q = 2^n - c$, a sufficient condition for one-step normalization is $c^2 + (2^n - 1) < 2(2^n - c)$. Rearranging this inequality yields
\begin{equation}
c^2 + 2c < 2^n + 1.
\label{eq:7}
\end{equation}
For integer \(c\), this condition is equivalent to \(c^2 + 2c \le 2^n\), which is consistent with the condition used in the folding-count analysis. Equivalently, it can also be written as
\begin{equation}
(c+1)^2 < 2^n + 2,
\label{eq:8}
\end{equation}
which can be rewritten as
\begin{equation}
c < \sqrt{2^n + 2} - 1.
\label{eq:9}
\end{equation}
Accordingly, the one-step normalization range can be approximated as
\begin{equation}
c \lesssim 2^{n/2}.
\label{eq:10}
\end{equation}
Therefore, for values of $c$ in this range, the result after two folding operations remains below $2q$, and the final modular reduction result can be obtained using a single conditional subtraction.

Next, we analyze the case in which final normalization is performed using two conditional subtractions. If the result $a(2)$ after two folding operations is smaller than $3q$, the final result can be obtained using two conditional subtractions, and thus we may impose the condition $a(2) < 3q$. Again using $a(2) \le c^2 + (2^n - 1)$ and $q = 2^n - c$, a sufficient condition for two-step normalization is $c^2 + (2^n - 1) < 3(2^n - c)$. Rearranging gives
\begin{equation}
c^2 + 3c < 2^{n+1} + 1.
\label{eq:11}
\end{equation}
Hence, when two-step normalization is allowed, the admissible range of $c$ becomes wider than that of the one-step normalization case, and can be approximated as
\begin{equation}
c \lesssim 2^{(n+1)/2}.
\label{eq:12}
\end{equation}

From the above analysis, the final normalization structure after two folding operations can be classified according to the range of the constant $c$ into a case using one conditional subtraction and a case using two conditional subtractions. Specifically, the condition \(c^2 + 2c < 2^n + 1\) gives a sufficient range in which the modular reduction result can be obtained using only one conditional subtraction. In the wider sufficient range satisfying \(c^2 + 3c < 2^{n+1} + 1\), normalization using up to two conditional subtractions is possible. Therefore, in the following section, we design the corresponding quantum Crandall reduction circuits based on this range analysis of $c$.
\section{Quantum circuit design for Crandall reduction}

In this section, we present the quantum circuit design for Crandall reduction based on the folding-count and normalization analyses in Section~2. The proposed circuit consists of folding and normalization stages specialized to the pseudo-Mersenne modulus \(q = 2^n - c\), and is built from quantum arithmetic components that exploit the fixed constants \(c\) and \(q\).

\subsection{Overview of the quantum Crandall reduction circuit}

\begin{figure}[h]
    \centering
    \includegraphics[height=0.5\textheight]{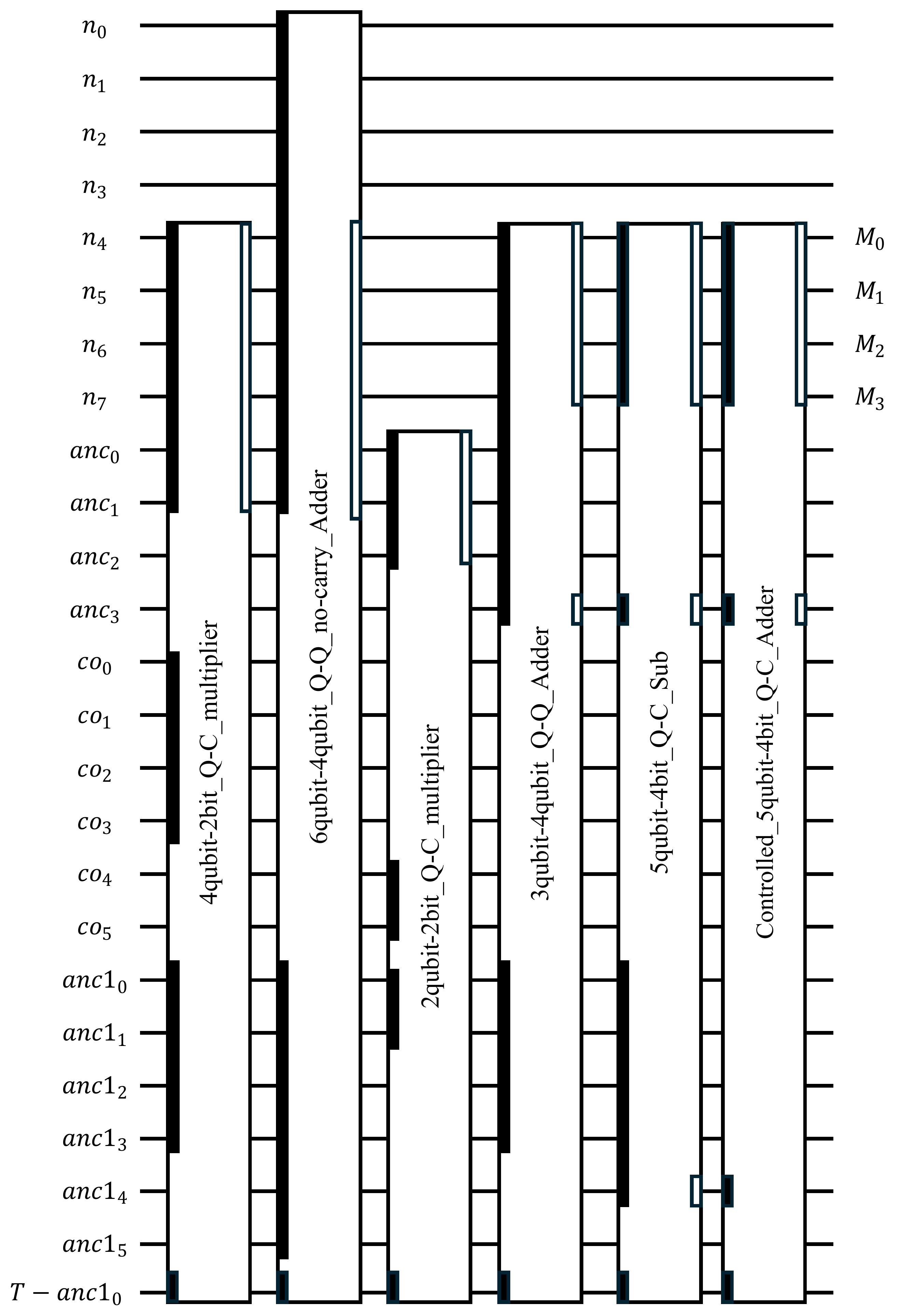}
    \caption{Overall structure of the quantum Crandall reduction circuit for \(n=4\) and \(c=3\). The folding stage consists of quantum-classical multiplication by \(c\) and quantum-quantum addition to accumulate the folded values. The normalization stage subtracts the fixed modulus \(q=2^n-c\) and conditionally restores it when the subtraction result is negative. After forward evaluation, the designated output register \(M\) contains \(a\bmod q\), while some ancillary work qubits retain input-dependent intermediate states until the inverse circuit is applied. The illustrated circuit corresponds to Crandall reduction-1 with one normalization block; Crandall reduction-2 applies the same normalization block twice.}
    \label{fig:overall}
\end{figure}

Each stage is implemented reversibly using quantum arithmetic components tailored to the fixed constants \(c\) and \(q\).

In the folding stage, the classical bit-shift operation is realized implicitly by register partitioning and wiring, and therefore does not require a separate quantum circuit. The remaining operations are multiplication by the fixed constant \(c\) and accumulation into the lower register. Since \(c\) is known at circuit design time, this multiplication is implemented as a quantum-classical multiplier rather than as a general quantum-quantum multiplier.

The normalization stage is implemented using a subtract-then-restore strategy. Instead of explicitly comparing the folded result with \(q\), the circuit first subtracts \(q=2^n-c\). If the subtraction result is negative, the original value is restored by conditionally adding \(q\) back. Since \(q\) is also a fixed classical constant, both the subtraction and the restoration addition are implemented as quantum-classical arithmetic circuits.

As an illustrative example, figure~\ref{fig:overall} shows the overall structure of the quantum Crandall reduction circuit for \(n=4\) and \(c=3\), corresponding to the one-step normalization variant. The two-step normalization variant applies the same normalization block once more. The following subsections focus on the two design principles that determine the logical resource costs: the temporary logical-AND-based arithmetic primitive and the quantum-classical specialization based on fixed classical bit patterns.

\subsection{Temporary logical-AND-based arithmetic primitive}

The arithmetic components used in the quantum Crandall reduction circuit rely on carry propagation, which is dominated by CCX gate operations. This dependence makes the choice of the CCX-level primitive a direct determinant of the \(T\)-count and \(T\)-depth of the overall circuit.

\begin{figure}[h]
    \centering
    \includegraphics[width=0.65\columnwidth]{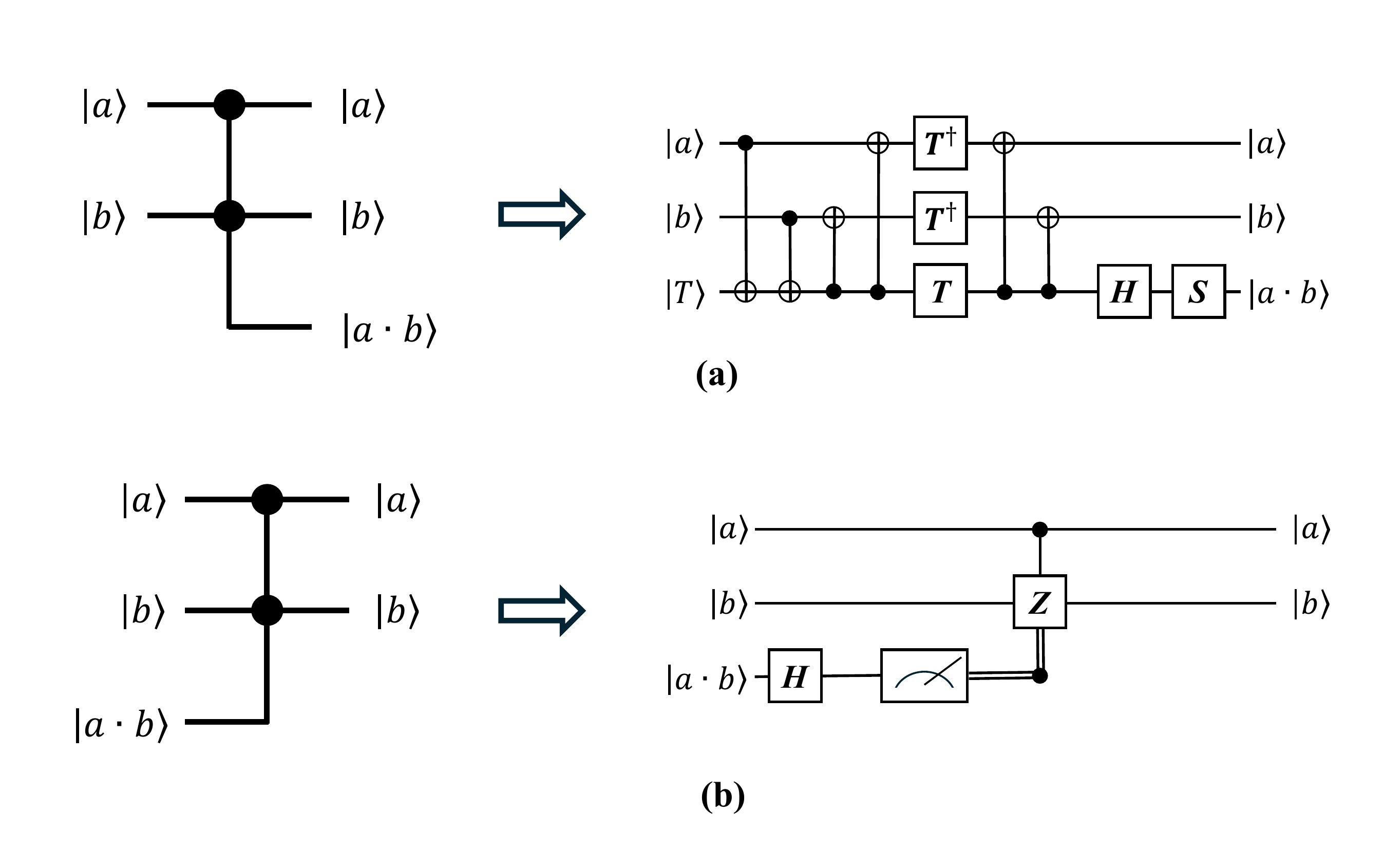}
    \caption{Temporary logical-AND quantum circuit. (a) Computation of the mapping
    \(|a\rangle|b\rangle|0\rangle \mapsto |a\rangle|b\rangle|a\cdot b\rangle\), which stores the logical-AND value used for carry or borrow propagation. (b) Measurement-and-fixup uncomputation of the stored logical-AND, which removes the temporary value without additional \(T\) gates.}
    \label{fig:temp_logical_and}
\end{figure}

In this work, we implement these CCX gate operations using Gidney’s temporary logical-AND construction rather than a conventional Toffoli gate decomposition~\cite{gidney2018halving}. Figure~\ref{fig:temp_logical_and} illustrates its computation and measurement-based uncomputation for carry and borrow propagation.

\begin{figure}[h]
    \centering
    \includegraphics[width=0.8\columnwidth]{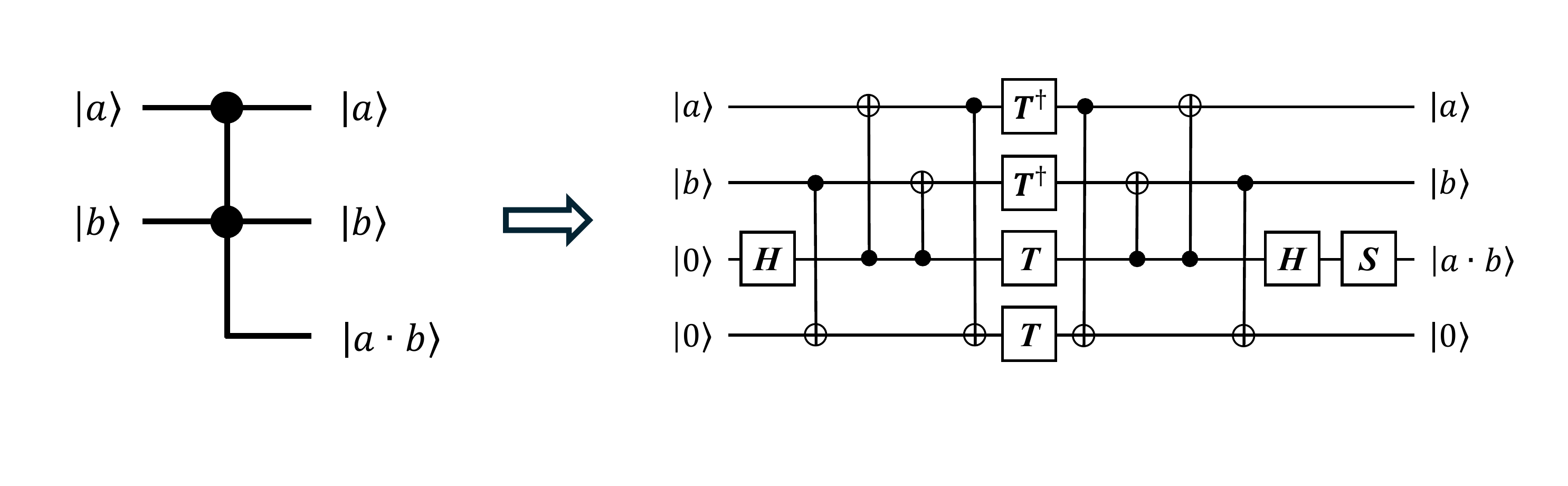}
    \caption{Temporary logical-AND computation circuit with \(T\)-depth 1. An additional temporary ancilla is used to parallelize the \(T/T^\dagger\) gates, reducing the \(T\)-depth from 2 to 1 while preserving the \(T\)-count of 4. The additional ancilla is restored to \(\lvert 0\rangle\) after the computation and can be reused in subsequent arithmetic operations.}
    \label{fig:1_depth}
\end{figure}

For the resource estimates in this paper, we use the \(T\)-depth-one variant of the temporary logical-AND computation shown in figure~\ref{fig:1_depth}. This construction preserves the \(T\)-count of 4 while reducing the \(T\)-depth from 2 to 1, and is therefore adopted as the CCX-level primitive throughout the resource analysis.

Based on this primitive, the resource costs of the component arithmetic circuits are obtained by counting the sequential CCX-level carry-generation blocks, each assigned \(T\)-count 4 and \(T\)-depth 1. The corresponding ripple-carry adder structure is provided in Appendix~A.

\subsection{Quantum-classical arithmetic with constant bit-pattern specialization}

In the quantum Crandall reduction circuit, several arithmetic operations involve one quantum operand and one fixed classical constant. In the folding stage, the upper quantum register is multiplied by the constant \(c\). In the normalization stage, the folded result is subtracted by \(q=2^n-c\), and \(q\) is conditionally added back when restoration is required. Since both \(c\) and \(q\) are known at circuit design time, these operations need not be implemented as general quantum-quantum arithmetic circuits.

\begin{figure}[h]
    \centering
    \includegraphics[width=0.55\columnwidth]{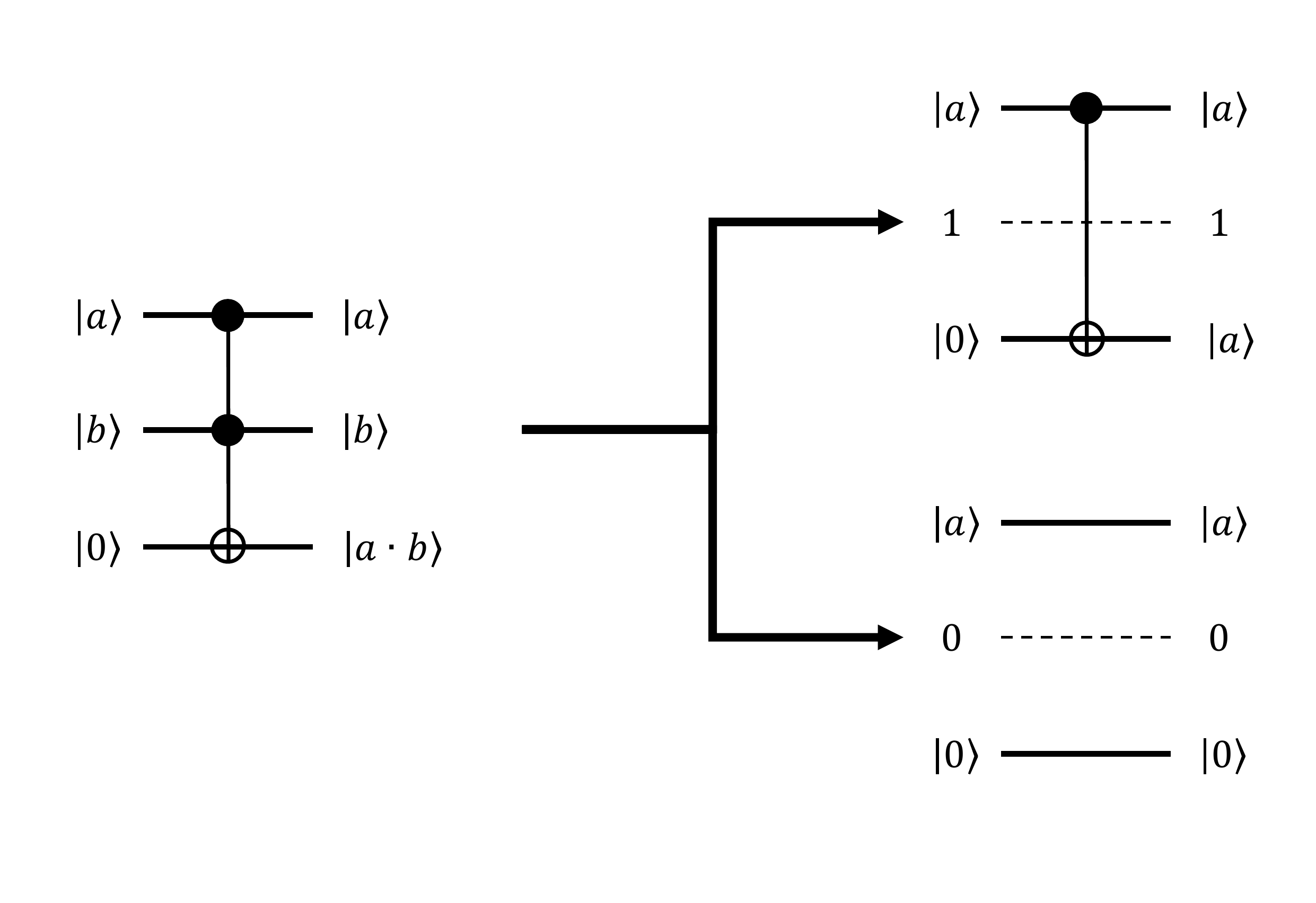}
    \caption{Simplification of a CCX gate operation when one control input is fixed as a classical bit and the target ancilla is initialized to \(|0\rangle\). If the classical bit is \(1\), the CCX gate reduces to a CX gate and the target becomes \(|a\rangle\). If the classical bit is \(0\), the operation is removed and the target remains \(|0\rangle\).}
    \label{fig:qc_bit_pattern_simplification}
\end{figure}

Figure~\ref{fig:qc_bit_pattern_simplification} illustrates the CCX gate simplification used to specialize the quantum-classical arithmetic circuits according to the fixed classical bit pattern.

This specialization is applied to the quantum-classical multiplier, subtractor, and controlled restoration adder. As a result, their resource costs depend on the Hamming weight and active bit positions of the fixed constants \(c\) and \(q\). Detailed circuit constructions are provided in Appendix~A, while the resulting resource formulas are presented in Section~4.
\section{Resource analysis and comparison}

In this section, we analyze the resource requirements of the quantum Crandall reduction circuit designed in Section~3 and compare them with those of existing Barrett reduction circuits. The proposed circuit consists of folding and normalization stages composed of a quantum-classical multiplier, quantum-quantum adder, quantum-classical subtractor, and controlled quantum-classical adder. The comparison targets are general Barrett reduction, folding Barrett reduction, and optimized folding Barrett reduction~\cite{zhang2025optimized}. The proposed circuits are classified as Crandall reduction-1 and Crandall reduction-2 according to the number of normalization steps. We compare the resource requirements of these circuits in terms of qubit count, \(T\)-count, and \(T\)-depth.

\subsection{Resource analysis of the quantum Crandall reduction circuit}

The total \(T\)-depth of the proposed circuit can be decomposed as
\begin{equation}
T_D^{\mathrm{CR}}
=
T_D^{\mathrm{fold}}
+
T_D^{\mathrm{norm}},
\label{eq:13}
\end{equation}
where \(T_D^{\mathrm{fold}}\) and \(T_D^{\mathrm{norm}}\) denote the \(T\)-depths of the folding and normalization stages, respectively.

The \(T\)-depth of the folding stage is
\begin{equation}
T_D^{\mathrm{fold}}
=
\left\lfloor \frac{n^2}{4} \right\rfloor
+
(n-1)\left\lfloor \frac{n}{2} \right\rfloor
+
n-2.
\label{eq:14}
\end{equation}

Each normalization step consists of a quantum-classical subtraction of \(q=2^n-c\), followed by a controlled restoration addition. Crandall reduction-1 performs this step once, giving
\begin{equation}
T_D^{\mathrm{norm1}}=4n+3,
\label{eq:15}
\end{equation}
whereas Crandall reduction-2 performs it twice, giving
\begin{equation}
T_D^{\mathrm{norm2}}=8n+6.
\label{eq:16}
\end{equation}

The overall \(T\)-depths of the two circuits are therefore
\begin{equation}
\begin{aligned}
T_D^{\mathrm{CR1}}
&=
\left\lfloor \frac{n^2}{4} \right\rfloor
+
(n-1)\left\lfloor \frac{n}{2} \right\rfloor
+
5n+1,\\
T_D^{\mathrm{CR2}}
&=
\left\lfloor \frac{n^2}{4} \right\rfloor
+
(n-1)\left\lfloor \frac{n}{2} \right\rfloor
+
9n+4.
\end{aligned}
\label{eq:17}
\end{equation}

For the \(T\)-count estimate, each CCX-level carry-generation block is implemented using the \(T\)-depth-one temporary logical-AND construction in figure~\ref{fig:1_depth}, which has a \(T\)-count of 4 and a \(T\)-depth of 1. The temporary logical-AND is uncomputed using measurement and fixup and therefore requires no additional \(T\) gates.

For a consistent comparison, all Barrett and Crandall reduction circuits are evaluated using the same temporary logical-AND construction, measurement-based uncomputation procedure, and qubit-counting convention. Therefore, the resource differences reported below arise from the circuit architectures rather than from different gate-decomposition assumptions.

\subsection{Quantum resource comparison}

Unless otherwise stated, the \(T\)-count and \(T\)-depth expressions for the proposed Crandall circuits represent worst-case estimates over the admissible range of \(c\). Specifically, the maximum Hamming weight derived in Appendix~\ref{app:t_depth_derivation} is used to bound the number of active partial products in quantum-classical multiplication. Consequently, instances with a sparser constant \(c\) may require fewer \(T\) gates and a smaller \(T\)-depth than the values reported here.

For compact presentation of the resource formulas in table~\ref{tab:resource_comparison}, we define
\begin{equation}
M(n)=\left\lceil \frac{n+1}{2} \right\rceil,
\qquad
F(n)=
\left\lfloor \frac{n^{2}}{4} \right\rfloor
+
(n-1)\left\lfloor \frac{n}{2} \right\rfloor.
\label{eq:resource_auxiliary_functions}
\end{equation}

\begin{table}[h]
\caption{Logical resource comparison of Barrett reduction circuits and the proposed Crandall reduction circuits.}
\centering
\renewcommand{\arraystretch}{1.35}

\begin{tabularx}{\textwidth}{
>{\raggedright\arraybackslash}X
>{\centering\arraybackslash}c
>{\centering\arraybackslash}c
>{\centering\arraybackslash}c
}
\hline
\textbf{Method} & \textbf{Qubit} & \textbf{\(T\)-count} & \textbf{\(T\)-depth} \\
\hline

General Barrett reduction
&
\(6n+6\)
&
\(8n^2+32n\)
&
\(2n^2+8n\)
\\

Folding Barrett reduction
&
\(6n+9\)
&
\(5n^2+44n\)
&
\(\dfrac{5}{4}n^2+11n\)
\\

Optimized folding Barrett reduction
&
\(6n+12\)
&
\(5n^2+38n+32\)
&
\(\dfrac{5}{4}n^2+\dfrac{19}{2}n+8\)
\\

Crandall reduction-1
&
\(4n+4M(n)+1\)
&
\(4F(n)+20n+4\)
&
\(F(n)+5n+1\)
\\

Crandall reduction-2
&
\(4n+4M(n)+2\)
&
\(4F(n)+36n+16\)
&
\(F(n)+9n+4\)
\\

\hline
\end{tabularx}
\label{tab:resource_comparison}
\end{table}

\begin{figure}[h]
\centering

\begin{subfigure}[t]{0.47\textwidth}
    \centering
    \includegraphics[width=\linewidth]{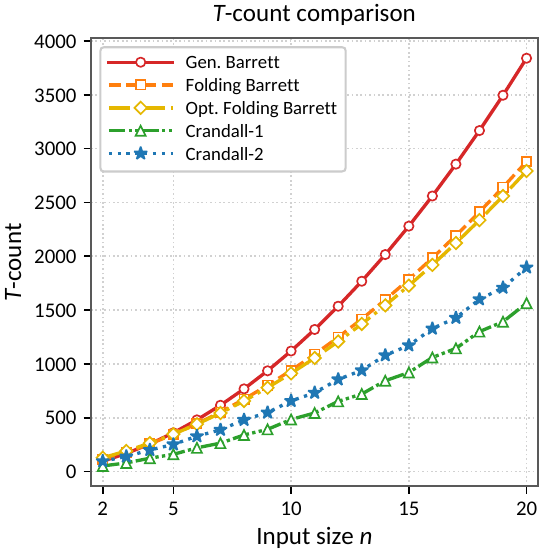}
    \caption{\(T\)-count comparison.}
    \label{fig:t_count_comparison}
\end{subfigure}
\hfill
\begin{subfigure}[t]{0.47\textwidth}
    \centering
    \includegraphics[width=\linewidth]{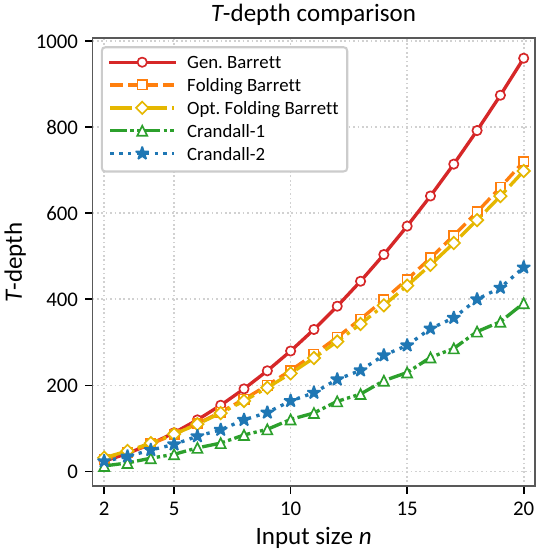}
    \caption{\(T\)-depth comparison.}
    \label{fig:t_depth_comparison}
\end{subfigure}

\caption{Logical \(T\)-count and \(T\)-depth comparison for Barrett reduction circuits and the proposed Crandall reduction circuits. Crandall reduction-1 achieves the lowest non-Clifford cost, while Crandall reduction-2 supports a wider range of \(c\) with moderate additional overhead.}
\label{fig:t_resource_comparison}
\end{figure}

Figure~\ref{fig:t_resource_comparison} shows the \(T\)-count and \(T\)-depth trends obtained from table~\ref{tab:resource_comparison}. Crandall reduction-1 gives the lowest \(T\)-count and \(T\)-depth, while Crandall reduction-2 remains more efficient than the Barrett reduction circuits over the plotted range despite its additional normalization step.

As a representative example, when \(n=10\), optimized folding Barrett reduction requires a \(T\)-count of \(912\) and a \(T\)-depth of \(228\). In contrast, Crandall reduction-1 requires a \(T\)-count of \(484\) and a \(T\)-depth of \(121\), corresponding to reductions of approximately \(46.9\%\) in both metrics. Crandall reduction-2 requires a \(T\)-count of \(656\) and a \(T\)-depth of \(164\), reducing both metrics by approximately \(28.1\%\) while supporting a wider admissible range of \(c\).

The resource advantage becomes more pronounced as \(n\) increases because the proposed circuits have smaller quadratic coefficients in both \(T\)-count and \(T\)-depth than the Barrett reduction circuits in table~\ref{tab:resource_comparison}. Thus, for pseudo-Mersenne moduli \(q=2^n-c\), their non-Clifford resource costs grow more slowly with the operand size.

The difference between Crandall reduction-1 and Crandall reduction-2 arises from the number of normalization steps. Crandall reduction-1 uses one normalization step and supports approximately \(c\lesssim 2^{n/2}\), whereas Crandall reduction-2 uses two normalization steps and extends the admissible range to approximately \(c\lesssim 2^{(n+1)/2}\). Consequently, Crandall reduction-2 incurs moderate additional \(T\)-count and \(T\)-depth but should be regarded as a range-extending variant rather than simply a higher-cost implementation.
\section{Surface-code-based resource analysis}
\label{sec:surface_code_resource}

To quantitatively analyze the fault-tolerant implementation cost of the proposed Crandall modular reduction circuits, we present a surface-code-based resource analysis model. Our analysis is based on the rotated planar surface code, and we assume that logical operations are performed using lattice-surgery operations~\cite{horsman2012surface}.

The resource analysis is carried out from two perspectives. First, for a fixed target circuit failure probability \(P_{\mathrm{fail}}\), we analyze how the difference in the \(T\)-depth among modular reduction circuits affects the allowable logical error rate and the required code distance. Second, by taking into account that \(T\) gates act as blocking operations in surface-code-based FTQC, we formulate the circuit execution time including both the \(T\)-state magic-state distillation time and the decoding time, and use this model to analyze and simulate the execution times of the modular reduction circuits.

\subsection{Code-distance analysis}
\label{subsec:failure_probability_code_distance}

To execute a circuit reliably, the logical error rate and the code distance must be chosen such that the total circuit failure probability remains below a target value. To quantify this requirement, we use the \(KQ\) formalism~\cite{steane2003overhead}, where \(K\) represents the number of logical qubits and \(Q\) represents the logical circuit size. The framework approximates the circuit failure probability by combining these quantities with the logical error rate, providing a simple model for fault-tolerant resource estimation. The \(KQ\) formalism used in this work covers circuit-level failures arising from effective logical gate errors under an independent stochastic error assumption. It does not explicitly account for coherent, correlated, leakage, or decoder-induced errors. These effects are mainly hardware- or decoder-dependent, rather than being intrinsic to the modular reduction circuits considered here. In this work, they are therefore assumed to be common across all algorithms and absorbed into an effective error-rate model. This allows the analysis to focus on the circuit-level differences among the modular reduction algorithms, such as logical-qubit count, \(T\)-depth, and the resulting surface-code distance.

In the \(KQ\) formalism, \(K\) denotes the number of logical qubits required to execute the circuit, while \(Q\) represents the scale of the logical circuit. In conventional resource analyses, \(Q\) is often defined as the total elementary operation count or the logical circuit depth. In this work, however, we interpret \(Q\) in relation to the \(T\)-depth, considering that \(T\) gates constitute the main bottleneck in both runtime and resource overhead in surface-code-based FTQC. Accordingly, the circuit failure probability is approximated as follows~\cite{ha2024resource}.

\begin{equation}
P_{\mathrm{fail}}
\approx
K D_T(n) \epsilon_L(n),
\label{eq:pfail}
\end{equation}

Here, \(n\) denotes half the number of bits in the input to the modular computation, \(K\) denotes the number of logical qubits required to execute the modular reduction circuit, \(D_T(n)\) denotes the \(T\)-depth of the circuit, and \(\epsilon_L(n)\) denotes the allowable logical error rate per logical \(T\)-layer or logical operation. Thus, for a fixed \(P_{\mathrm{fail}}\), a smaller \(K D_T(n)\) allows a higher \(\epsilon_L(n)\) and consequently a smaller required code distance.

In the rotated surface code, the logical error rate decreases exponentially with the physical error rate and the code distance. We use the following surface-code logical error model~\cite{fowler2019low}.

\begin{equation}
\epsilon_L
=
0.1
\left(
100\epsilon_p
\right)^{(d+1)/2},
\label{eq:logical_error_model}
\end{equation}

Here, \(\epsilon_p\) denotes the physical gate error rate, and \(d\) denotes the surface-code distance. By rearranging the above expression with respect to \(d\), the code distance required to satisfy the target logical error rate \(\epsilon_L(n)\) can be calculated as follows.

\begin{equation}
d(n)
=
2
\left\lceil
\frac{
\log\left(10\epsilon_L(n)\right)
}{
\log\left(100\epsilon_p\right)
}
\right\rceil
-1.
\label{eq:code_distance}
\end{equation}

Using the relation between equations~\eqref{eq:pfail} and~\eqref{eq:code_distance}, we compare the code-distance requirements of general Barrett reduction, optimized folding Barrett reduction, Crandall reduction-1, and Crandall reduction-2. For each circuit, the \(T\)-depth \(D_T(n)\) obtained from the logical-circuit analysis is substituted into the above expression, and the required code distance \(d(n)\) is calculated under the same values of \(P_{\mathrm{fail}}\), \(K\), and \(\epsilon_p\). Table~\ref{tab:code_distance} summarizes the resource analysis of each circuit for \(P_{\mathrm{fail}}=0.1\)~\cite{ogorman2017quantum} and \(\epsilon_p=10^{-3}\)~\cite{google2025quantum}.

\begin{table}[t]
\caption{Required surface-code distance \(d(n)\) for each modular reduction circuit. The target failure probability \(P_{\mathrm{fail}}=0.1\) follows the resource-estimation setting of Ref.~\cite{ogorman2017quantum}, and the physical error rate is taken as \(\epsilon_p=10^{-3}\) based on the experimental results of Ref.~\cite{google2025quantum}.}
\centering
\renewcommand{\arraystretch}{1.4}
\setlength{\extrarowheight}{2pt}
\setlength{\tabcolsep}{5pt}

\begin{tabularx}{\textwidth}{
    >{\raggedright\arraybackslash}p{0.28\textwidth}
    >{\centering\arraybackslash}X
}
\hline
\textbf{Method} & \textbf{Code distance \(d(n)\)} \\
\hline
\noalign{\vskip 1.2ex}

General Barrett reduction
&
\(\displaystyle
2\left\lceil
\log_{10}\!\left(
12n^3+60n^2+48n
\right)
\right\rceil-1
\)
\\[2.5ex]

Optimized folding Barrett reduction
&
\(\displaystyle
2\left\lceil
\log_{10}\!\left(
\frac{15n^3+144n^2+324n+192}{2}
\right)
\right\rceil-1
\)
\\[2.5ex]

Crandall reduction-1
&
\(\displaystyle
\begin{cases}
2\left\lceil
\log_{10}\!\left(
\frac{18n^3+93n^2+60n+9}{4}
\right)
\right\rceil-1,
& n\ \mathrm{odd}, \\[2.6ex]
2\left\lceil
\log_{10}\!\left(
\frac{18n^3+123n^2+114n+20}{4}
\right)
\right\rceil-1,
& n\ \mathrm{even}.
\end{cases}
\)
\\[6ex]

Crandall reduction-2
&
\(\displaystyle
\begin{cases}
2\left\lceil
\log_{10}\!\left(
\frac{9n^3+102n^2+115n+34}{2}
\right)
\right\rceil-1,
& n\ \mathrm{odd}, \\[2.6ex]
2\left\lceil
\log_{10}\!\left(
\frac{9n^3+111n^2+150n+48}{2}
\right)
\right\rceil-1,
& n\ \mathrm{even}.
\end{cases}
\)
\\

\noalign{\vskip 0.8ex}
\hline
\end{tabularx}
\label{tab:code_distance}
\end{table}

Since the quantum Crandall reduction circuit shown in figure~\ref{fig:overall} was analyzed for \(n=4\), we also calculated the required code distance for \(n=4\) under each target algorithm failure probability. The results are presented in table~\ref{tab:appendix_code_distance}.

\begin{table}[h]
\caption{Required surface-code distance for modular reduction circuits at \(n=4\) under different target circuit failure probabilities \(P_{\mathrm{fail}}\).}
\centering
\renewcommand{\arraystretch}{1.35}
\setlength{\tabcolsep}{4pt}

\begin{tabularx}{\textwidth}{
    >{\centering\arraybackslash}c |
    >{\centering\arraybackslash}X
    >{\centering\arraybackslash}X
    >{\centering\arraybackslash}X
    >{\centering\arraybackslash}X
}
\hline
\textbf{\(P_{\mathrm{fail}}\)}
&
\textbf{General Barrett reduction}
&
\textbf{Optimized folding Barrett reduction}
&
\textbf{Crandall reduction-1}
&
\textbf{Crandall reduction-2}
\\
\hline
\noalign{\vskip 0.1ex}

\(10^{-6}\) & \(17\) & \(17\) & \(15\) & \(17\) \\[0.1ex]

\(10^{-5}\) & \(15\) & \(15\) & \(13\) & \(15\) \\[0.1ex]

\(10^{-4}\) & \(13\) & \(13\) & \(11\) & \(13\) \\[0.1ex]

\(10^{-3}\) & \(11\) & \(11\) & \(9\)  & \(11\) \\[0.1ex]

\(10^{-2}\) & \(9\)  & \(9\)  & \(7\)  & \(9\) \\[0.1ex]

\(10^{-1}\) & \(7\)  & \(7\)  & \(5\)  & \(7\) \\

\noalign{\vskip 0.1ex}
\hline
\end{tabularx}
\label{tab:appendix_code_distance}
\end{table}

Table~\ref{tab:appendix_code_distance} summarizes the required surface-code distances for the modular reduction circuits under different target circuit failure probabilities. The results show that Crandall reduction-1 consistently requires a smaller code distance than the other circuits under stricter reliability requirements, indicating a lower physical-qubit overhead in fault-tolerant implementation.

\subsection{Magic-state factory time model}
\label{subsec:magic_state_factory_time_model}

In the surface code, a \(T\) gate is typically implemented through magic-state injection or magic-state teleportation, and magic-state distillation is required to obtain a high-fidelity \(T\) state at the fault-tolerant level~\cite{litinski2019game,haah2017magic}. In this work, we model the latency of the magic-state factory required for executing a \(T\) gate as

\begin{equation}
T_{\mathrm{MSF}}(n)
=
6.5\cdot d(n)\cdot c_t,
\label{eq:t_msf}
\end{equation}

where \(d(n)\) is the code distance calculated in the previous subsection and \(c_t\) denotes one surface-code cycle time. The constant factor \(6.5\) represents the distillation latency converted into units of code distance, based on the magic-state factory architecture of Fowler and Gidney~\cite{fowler2019low}.

Accordingly, the total magic-state factory execution time of the circuit is given by the product of the magic-state factory latency per \(T\)-layer and the \(T\)-depth. Thus,

\begin{equation}
t_{\mathrm{MSF}}(n)
=
6.5\cdot d(n)\cdot c_t\cdot D_T(n).
\label{eq:t_msf_total}
\end{equation}

\subsection{Proposed circuit runtime model}
\label{subsec:proposed_runtime_model}

A \(T\) gate is a blocking operation, so it may be necessary to wait until the decoder results come out~\cite{fowler2012surface}. Although real-time processing is desirable, the high syndrome generation rate makes it challenging to process all syndromes in a timely manner. When the syndrome generation rate is denoted by \(r_s\) and the processing throughput of the classical decoder by \(r_{\mathrm{proc}}\), the ratio between these two quantities determines whether the decoder backlog grows. If \(r_{\mathrm{proc}} > r_s\), the decoder can keep up with the generated syndrome data, and the backlog decreases or remains below a certain level. Conversely, if \(r_s > r_{\mathrm{proc}}\), unprocessed syndrome data accumulate, resulting in an increasing decoder backlog.

At the execution time of the \(k\)-th \(T\) gate, the accumulated decoder backlog is denoted by \(\Delta_k\), with \(\Delta_1\) representing the initial backlog at the time of the first \(T\) gate. The backlog can then be expressed as
\(\Delta_k
=
f^{k-1}\cdot\Delta_1\),
where \(f\) denotes the ratio of the syndrome generation rate to the decoder processing throughput~\cite{terhal2015quantum}. To capture the first-order impact of decoder throughput limitations on the execution time, we adopt a simplified exponential backlog model. This model is intended to provide a coarse-grained analytical estimate of decoder-limited execution rather than an exact prediction of hardware runtime. If \(f>1\), the backlog increases exponentially as the \(T\)-depth increases. Conversely, if \(f\leq1\), the decoder can keep up with syndrome generation, and the backlog does not grow or becomes stabilized. An additional analysis of the cumulative decoder burden based on the proposed runtime model is presented in Appendix~\ref{subsec:decoder_burden}.

The decoder runtime required to process the \(k\)-th backlog is given by the backlog size divided by the decoder processing throughput.

\begin{equation}
T_{\mathrm{dec}}(\Delta_k)
=
\frac{\Delta_k}{r_{\mathrm{proc}}}.
\label{eq:decoding_time}
\end{equation}

Therefore, the time required to process the \(T\) gate at the \(T\)-depth is given by

\begin{equation}
t_{\mathrm{backlog}}(n)
=
\frac{\Delta_{D_T(n)}}{r_{\mathrm{proc}}}
=
\frac{f^{D_T(n)-1}\cdot\Delta_1(n)}{r_{\mathrm{proc}}}.
\label{eq:backlog_latency}
\end{equation}

By incorporating the latency caused by the decoder backlog into the overall execution time analysis, the total circuit execution time can be expressed as the sum of the Clifford gate execution time, the \(T\)-state magic-state factory execution time, and the decoder backlog time. Therefore, the total runtime is expressed as

\begin{equation}
t_{\mathrm{alg}}(n)
=
t_{\mathrm{Clifford}}
+
6.5\cdot d(n)\cdot c_t\cdot D_T(n)
+
\frac{f^{D_T(n)-1}\cdot\Delta_1(n)}{r_{\mathrm{proc}}}.
\label{eq:runtime_full}
\end{equation}

Since the execution time of Clifford gates is much smaller than the magic-state factory latency~\cite{ogorman2017quantum,haah2017magic} and the decoder waiting time in the considered FTQC setting~\cite{terhal2015quantum,tan2023scalable}, the total circuit execution time can be approximated as

\begin{equation}
t_{\mathrm{alg}}(n)
\approx
6.5\cdot d(n)\cdot c_t\cdot D_T(n)
+
\frac{f^{D_T(n)-1}\cdot\Delta_1(n)}{r_{\mathrm{proc}}}.
\label{eq:runtime_approx}
\end{equation}

\subsection{Simulation}
\label{subsec:simulation}

We use the proposed circuit runtime model to compare the execution times of modular reduction circuits in a fault-tolerant quantum computing environment. The circuits compared are general Barrett reduction, optimized folding Barrett reduction, Crandall reduction-1, and Crandall reduction-2. Since optimized folding Barrett reduction is an optimized version of folding Barrett reduction, only optimized folding Barrett reduction is included in this simulation.

The gate configuration before the first \(T\) gate differs among the circuits. In general Barrett reduction and optimized folding Barrett reduction~\cite{zhang2025optimized}, \(2n+4\) CNOT gates and one Hadamard gate are executed before the first \(T\) gate. In contrast, in Crandall reduction-1 and Crandall reduction-2, \(n+4\) CNOT gates and one Hadamard gate are executed before the first \(T\) gate.

We assume that surface-code operations are performed based on lattice surgery~\cite{horsman2012surface}. The number of syndromes generated during one CNOT operation using lattice surgery is \(6d^3-6d\). We also assume that the Hadamard gate is implemented using lattice surgery. Also, the number of syndromes generated during a Hadamard operation using lattice surgery is \(4d^3-4d\). Based on this, by substituting the code distance \(d(n)\) obtained for each circuit in table~\ref{tab:code_distance}, the initial backlog of general Barrett reduction and optimized folding Barrett reduction is \((12n+28)(d(n)^3-d(n))\), and the initial backlog of Crandall reduction-1 and Crandall reduction-2 is \((6n+28)(d(n)^3-d(n))\).

\begin{figure}[h]
    \centering

    \begin{subfigure}[t]{0.49\textwidth}
        \centering
        \includegraphics[width=\linewidth]
        {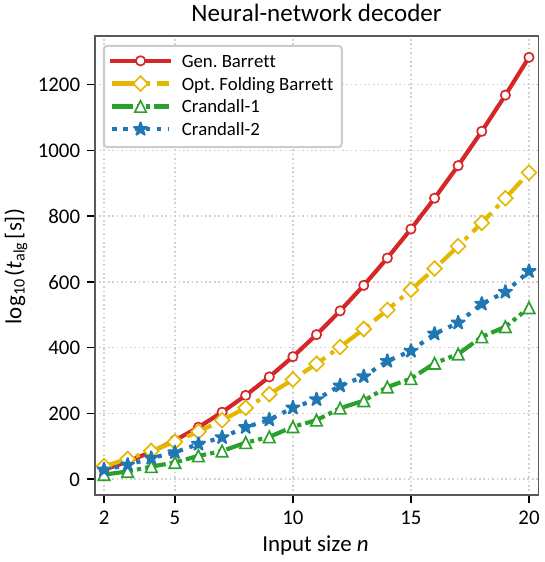}
        \caption{Neural-network decoder.}
        \label{fig:runtime_nn_decoder}
    \end{subfigure}
    \hfill
    \begin{subfigure}[t]{0.49\textwidth}
        \centering
        \includegraphics[width=\linewidth]
        {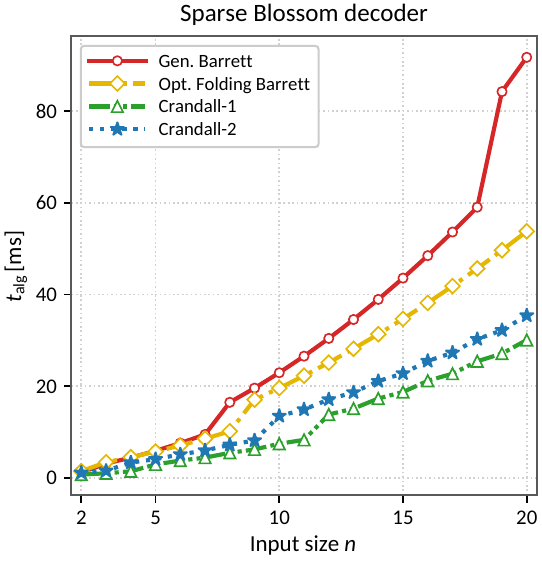}
        \caption{Sparse Blossom decoder.}
        \label{fig:runtime_sparse_blossom}
    \end{subfigure}

    \caption{Comparison of execution times for modular reduction circuits in a surface-code-based FTQC environment. The execution time is calculated using the runtime model in equation~\eqref{eq:runtime_approx}, where \(D_T(n)\), \(d(n)\), and \(\Delta_1(n)\) are obtained from the logical resource estimates and code-distance conditions of each circuit. (a) Under the neural-network decoder setting \((f=21.8,\ r_{\mathrm{proc}}=10^6~\mathrm{bits/s})\)~\cite{google2025quantum}, the decoder backlog grows rapidly with \(T\)-depth; therefore, the result is plotted on a \(\log_{10}(t_{\mathrm{alg}}/\mathrm{s})\) scale. (b) Under the Sparse Blossom decoder setting \((f=1,\ r_{\mathrm{proc}}=2.18\times10^7~\mathrm{bits/s})\)~\cite{google2025quantum}, the backlog is not exponentially amplified, and the execution time is presented on a millisecond scale. For both decoder settings, Crandall reduction-1 achieves the shortest execution time, followed by Crandall reduction-2.}
    \label{fig:runtime_comparison}
\end{figure}

The syndrome generation rate and decoder processing throughput used in the circuit execution time model in equation~\eqref{eq:runtime_approx} are set based on the \(d=5\) condition reported in Ref.~\cite{google2025quantum}. When the cycle time is \(1.1~\mu\mathrm{s}\), and 24 syndrome bits are generated per cycle for \(d=5\), the syndrome generation rate is \(2.18\times10^7~\mathrm{bits/s}\). The decoder processing throughput is \(10^6~\mathrm{bits/s}\) for the neural-network decoder and \(2.18\times10^7~\mathrm{bits/s}\) for the Sparse Blossom decoder. Therefore, the backlog growth factor \(f=r_s/r_{\mathrm{proc}}\) is 21.8 and 1, respectively. The circuit runtimes for the neural-network decoder and the Sparse Blossom decoder, calculated using equation~\eqref{eq:runtime_approx}, are shown in figure~\ref{fig:runtime_comparison}.

For both decoders, Crandall reduction-1 achieves the shortest execution time, followed by Crandall reduction-2. The relative ordering between general Barrett reduction and optimized folding Barrett reduction depends on the input size, with optimized folding Barrett reduction becoming more efficient for \(n\geq5\). Under the neural-network decoder condition, the execution time was compared on the \(\log_{10}(t_{\mathrm{alg}}[\mathrm{s}])\) scale. At \(n=20\), the corresponding values for general Barrett reduction, optimized folding Barrett reduction, Crandall reduction-1, and Crandall reduction-2 were \(1283.13\), \(932.19\), \(521.03\), and \(632.12\), respectively. Under the proposed runtime model, Crandall reduction-1 and Crandall reduction-2 reduce the estimated execution time by several orders of magnitude relative to the two Barrett-based circuits. Under the Sparse Blossom decoder condition, at \(n=20\), the execution times of general Barrett reduction, optimized folding Barrett reduction, Crandall reduction-1, and Crandall reduction-2 were \(91.73~\mathrm{ms}\), \(53.77~\mathrm{ms}\), \(30.05~\mathrm{ms}\), and \(35.39~\mathrm{ms}\), respectively. Crandall reduction-1 reduced the execution time by \(67.2\%\) and \(44.1\%\) compared with general Barrett reduction and optimized folding Barrett reduction, respectively, while Crandall reduction-2 reduced it by \(61.4\%\) and \(34.2\%\), respectively.

In the execution-time graph obtained using the Sparse Blossom decoder, discontinuous increases in the execution time appear at some input sizes. This is because, when \(f=1\), the backlog is not exponentially amplified with the \(T\)-depth, and the points at which the code distance \(d(n)\) changes are directly reflected as jumps in the execution time. Therefore, when \(d(n)\) increases from \(5\) to \(7\), the backlog term increases by a factor of approximately \(2.14\), making the execution time appear to jump upward at specific values of \(n\). For example, the large jumps in general Barrett reduction occur near \(n=8\), where \(d(n)\) increases from \(7\) to \(9\), and near \(n=19\), where \(d(n)\) increases from \(9\) to \(11\). In Crandall reduction-1, \(d(n)\) increases from \(7\) to \(9\) near \(n=12\), producing a similar increase in the execution time around that point.
\section{Conclusion}

This work demonstrates that exploiting the algebraic structure of a modulus can improve quantum modular reduction from logical-circuit design through fault-tolerant implementation, as shown here for pseudo-Mersenne moduli. To provide a structure-aware alternative to Barrett reduction for moduli of the form \(q=2^n-c\), we adapted Crandall folding to reversible computation and developed two circuit variants: Crandall reduction-1 with one-step normalization and Crandall reduction-2 with two-step normalization.

At the logical level, both proposed circuits achieve lower \(T\)-count and \(T\)-depth than the Barrett reduction circuits considered. For \(n=10\), Crandall reduction-1 reduces both metrics by approximately \(46.9\%\) relative to optimized folding Barrett reduction. Under the same condition, Crandall reduction-2 reduces both metrics by approximately \(28.1\%\), while extending the admissible range from \(1\leq c\leq31\) to \(1\leq c\leq43\).

To account for logical-error accumulation arising from repeated non-Clifford operations in modular-reduction circuits, we mapped the proposed circuits onto a surface-code-based fault-tolerant architecture and analyzed the required code distance and execution time. At \(n=4\) and \(P_{\mathrm{fail}}=10^{-6}\), the required code distance for Crandall reduction-1 decreases from \(17\) to \(15\) relative to optimized folding Barrett reduction. Under the Sparse Blossom decoder setting at \(n=20\), the estimated runtime of Crandall reduction-1 is \(23.72~\mathrm{ms}\) shorter than that of optimized folding Barrett reduction. These results show that the structural advantage obtained at the logical level persists after surface-code mapping and can reduce the implementation cost of fault-tolerant quantum arithmetic. Although the present construction is specialized for pseudo-Mersenne moduli, it provides a concrete example of the benefit of structure-aware circuit design. Future work will extend this principle to other special-form moduli and evaluate the proposed circuits within complete quantum algorithms.

\ack{This research was supported by the Korea Institute of Science and Technology Information (KISTI) (No.~(KISTI)K26L1M3C5-01). This work was also supported by the Development of Core Technologies for the Nationwide Transition to Post-Quantum Cryptography (R\&D) program funded by the Ministry of Science and ICT, Republic of Korea, through the project ``Development of an Inter-Domain Quantum Security System Combining PQC and QKD'' (No.~N26NM042-26).}

\appendix

\section{Component-level arithmetic circuits}
\label{app:component_circuits}

This appendix provides the component-level circuits used in the quantum Crandall reduction circuit. The implementations follow the temporary logical-AND primitive and fixed bit-pattern specialization introduced in Sections~3.2 and~3.3. Detailed structures are given for the ripple-carry adder, quantum-classical multiplier, quantum-classical subtractor, and controlled restoration adder.

\subsection{Ripple-carry adder based on temporary logical-AND}
\label{app:ripple_carry_adder}

\begin{figure}[h]
    \centering
    \includegraphics[width=0.85\columnwidth]{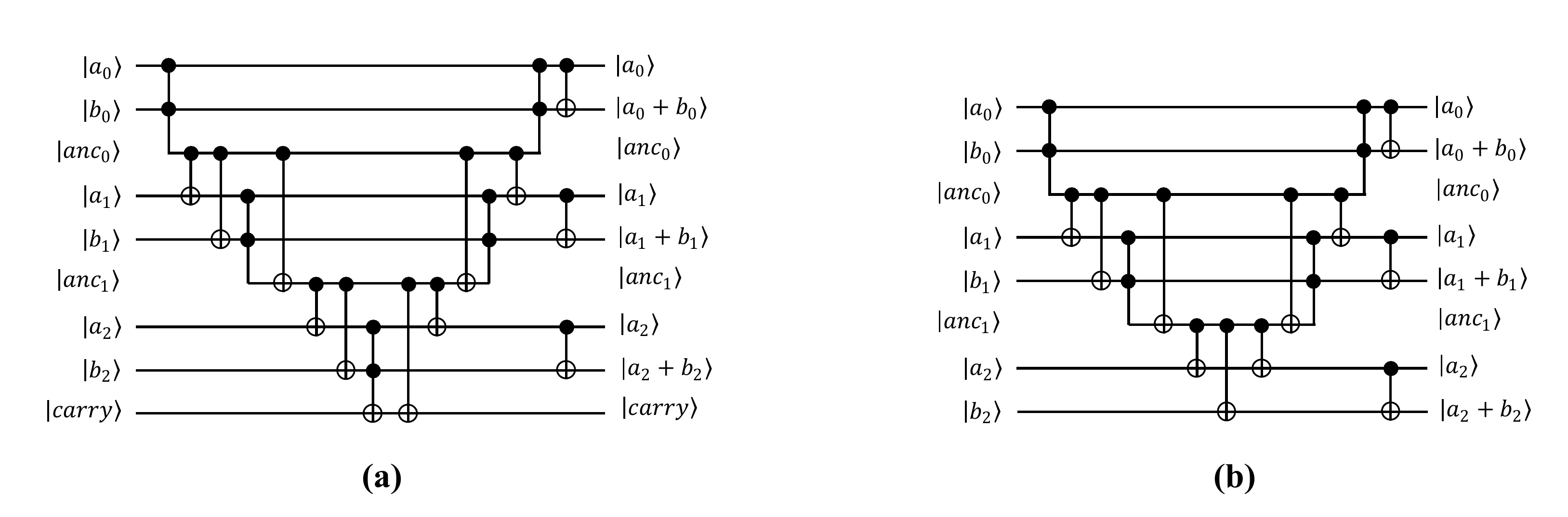}
    \caption{Quantum circuits for ripple-carry adders based on the temporary logical-AND primitive. (a) Standard ripple-carry adder including the final carry-out. (b) Modified ripple-carry adder without the final carry-out.}
    \label{fig:app_ripple_adder}
\end{figure}

Figure~\ref{fig:app_ripple_adder} shows the two ripple-carry adder structures used in the folding stage. Both propagate the carry from the least significant bit to the most significant bit and uncompute the temporary carry values in reverse order. The standard form includes the final carry-out, whereas the no-carry form omits the last carry-generation block when the final carry bit is not required.

For the resource analysis, each sequential carry-generation block is assigned a \(T\)-count of 4 and a \(T\)-depth of 1. The resource cost of each adder is therefore determined by the number of carry-generation blocks in the corresponding structure.

\subsection{Quantum-classical multiplier}
\label{app:qc_multiplier}

The folding stage multiplies a quantum register by the fixed classical constant \(c\) using shifted partial-product additions associated with the 1-bits of \(c\).

\begin{figure}[h]
    \centering
    \includegraphics[width=0.6\columnwidth]{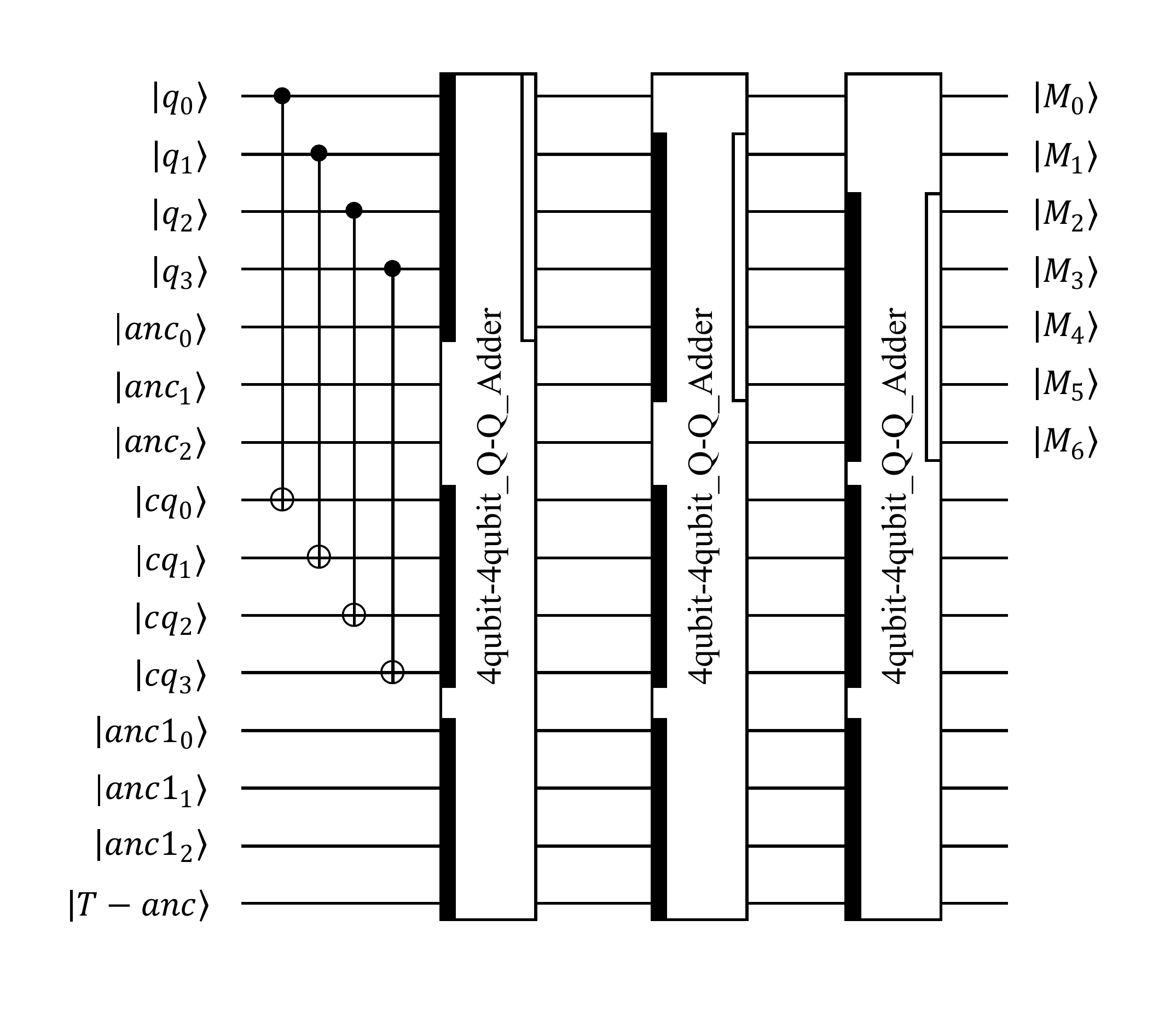}
    \caption{Quantum-classical multiplier for a 4-qubit quantum operand and the 3-bit classical constant \(111_2\). Adders associated with zero bits of the constant are omitted in the specialized circuit.}
    \label{fig:app_qc_multiplier}
\end{figure}

Figure~\ref{fig:app_qc_multiplier} shows an example for a 4-qubit quantum operand and the 3-bit constant \(111_2\), for which all shifted partial-product additions are present. For a general constant \(c\), adders associated with zero bits are omitted.

Accordingly, the multiplier resource cost is determined by the Hamming weight and active bit positions of \(c\).

\subsection{Quantum-classical subtractor and controlled restoration adder}
\label{app:qc_subtractor_restoration}

The normalization block consists of subtraction by the fixed modulus \(q=2^n-c\), followed by a controlled restoration addition.

\begin{figure}[h]
    \centering
    \includegraphics[width=0.9\columnwidth]{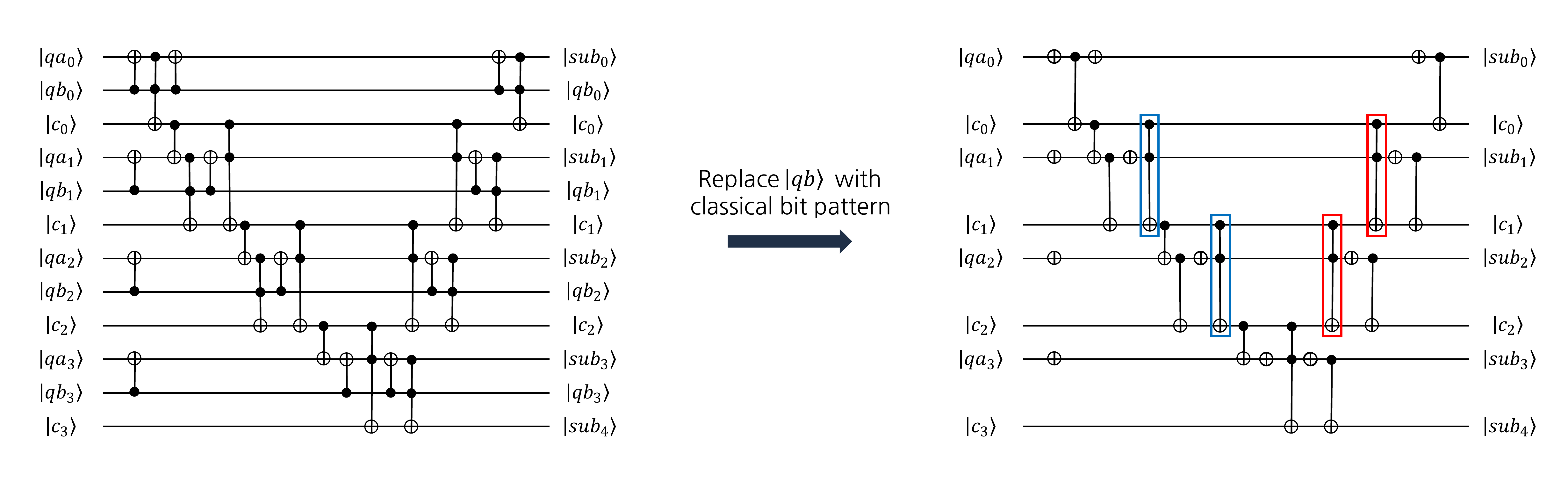}
    \caption{Quantum-classical subtraction circuit. The left-hand circuit shows the quantum-quantum subtractor based on the inverse temporary logical-AND-based adder, and the right-hand circuit shows its quantum-classical form after replacing one operand with the bit pattern of the classical constant \(q\).}
    \label{fig:app_qc_subtractor}
\end{figure}

Figure~\ref{fig:app_qc_subtractor} shows the subtractor used in this block. The left-hand circuit is the quantum-quantum subtractor obtained from the inverse ripple-carry adder, and the right-hand circuit is its quantum-classical form specialized for the bit pattern of \(q\).

The sign information is stored in the most significant qubit of the subtraction result. A negative result sets this qubit, which controls the restoration addition of \(q\).

\begin{figure}[h]
    \centering
    \includegraphics[width=0.9\columnwidth]{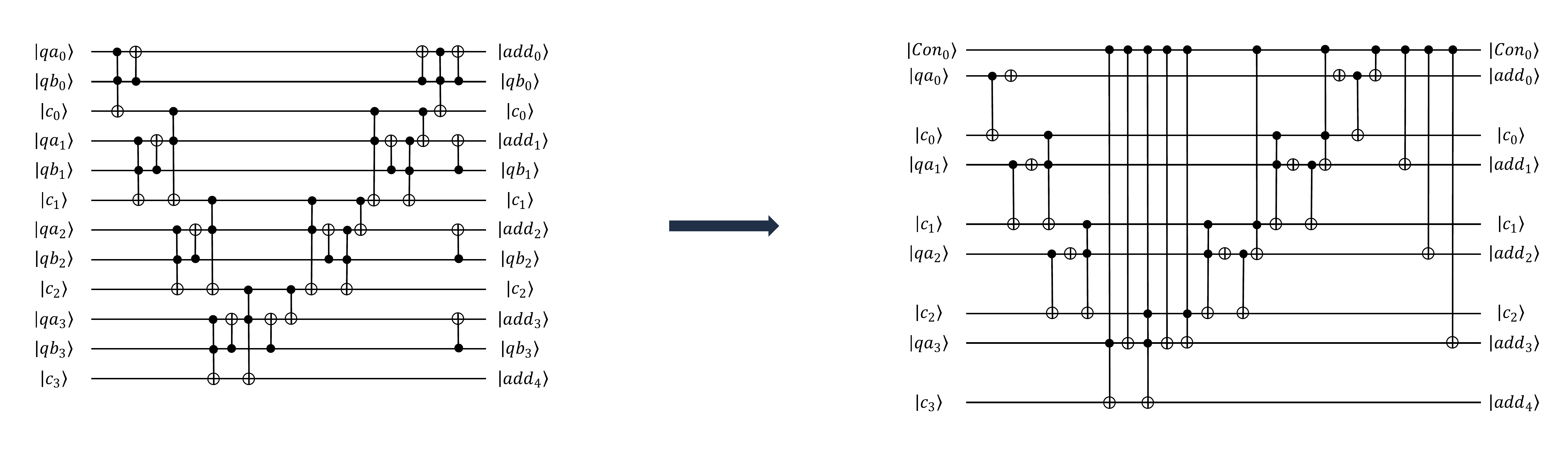}
    \caption{Controlled quantum-classical adder circuit used for the restoration step in normalization. The control qubit determines whether the fixed classical constant \(q=2^n-c\) is added back. The bit pattern of \(q\) is embedded into the circuit to remove unnecessary gates.}
    \label{fig:app_controlled_qc_adder}
\end{figure}

Figure~\ref{fig:app_controlled_qc_adder} shows the controlled quantum-classical adder used for this restoration step.

Crandall reduction-1 applies this normalization block once, whereas Crandall reduction-2 applies it twice. The sign qubit is included as part of the normalization workspace in the resource estimate.


\section{Derivation of the \texorpdfstring{\(T\)}{T}-depth formulas}
\label{app:t_depth_derivation}

This appendix derives the worst-case \(T\)-depth and \(T\)-count expressions used in Section~4 by accounting for the active bit pattern of the fixed classical constant \(c\).

Let \(\mathrm{wt}(c)\) denote the Hamming weight of \(c\), and let \(\mathcal{C}\) be the set of admissible constants satisfying the applicable normalization condition. The maximum number of active bits is defined as
\begin{equation}
k = \max_{c\in\mathcal{C}} \mathrm{wt}(c).
\end{equation}
Since the smallest positive integer with \(r\) ones in its binary representation is
\begin{equation}
2^r - 1,
\end{equation}
and the left-hand side of each admissibility condition increases with \(c\), a Hamming weight of at least \(r\) is possible only if \(2^r-1\) satisfies the corresponding condition.

For one-step normalization, the admissibility condition for integer \(c\) is
\begin{equation}
c^2+2c \leq 2^n.
\end{equation}
Substituting \(c=2^r-1\) gives
\begin{equation}
\begin{aligned}
c^2+2c
&=
(2^r-1)^2+2(2^r-1) \\
&=
2^{2r}-1.
\end{aligned}
\end{equation}
For \(r=\left\lfloor n/2 \right\rfloor\), this condition is satisfied because \(2^{2r}-1 \leq 2^n\). By contrast, the smallest integer with \(r+1\) ones is \(2^{r+1}-1\), for which the left-hand side becomes \(2^{2r+2}-1>2^n\). Hence, for one-step normalization,
\begin{equation}
k=\left\lfloor \frac{n}{2} \right\rfloor.
\end{equation}

For two-step normalization, the admissibility condition is
\begin{equation}
c^2+3c \leq 2^{n+1}.
\end{equation}
Substitution of \(c=2^r-1\) yields
\begin{equation}
\begin{aligned}
c^2+3c
&=
(2^r-1)^2+3(2^r-1) \\
&=
2^{2r}+2^r-2.
\end{aligned}
\end{equation}
For \(r=\left\lfloor n/2 \right\rfloor\), this value is at most \(2^{n+1}\). Substitution of \(2^{r+1}-1\), the smallest integer with \(r+1\) ones, gives \(2^{2r+2}+2^{r+1}-2\), which violates the condition. Therefore, the worst-case Hamming weight for both normalization variants is
\begin{equation}
k=\left\lfloor \frac{n}{2} \right\rfloor.
\end{equation}

The accumulation depth also depends on the positions of the active bits. We therefore use the effective active-bit range
\begin{equation}
m=\left\lceil \frac{n}{2} \right\rceil.
\end{equation}
Although some admissible constants under two-step normalization may have a most significant bit above this range, the admissibility condition prevents such constants from simultaneously attaining the maximum Hamming weight \(k\). Their increased shift range is therefore offset by a smaller number of active partial products. Consequently, the worst-case sequential accumulation cost is bounded by \(k\) active bits within the effective range \(m\). Accordingly,
\begin{equation}
mk
=
\left\lceil \frac{n}{2} \right\rceil
\left\lfloor \frac{n}{2} \right\rfloor
=
\left\lfloor \frac{n^2}{4} \right\rfloor.
\end{equation}

We next derive the \(T\)-depths of the two folding operations. In the first quantum-classical multiplication, the first partial product is used as the initial value, and the remaining \(k-1\) partial products are accumulated over an \(n\)-qubit register. Its \(T\)-depth is therefore
\begin{equation}
n(k-1).
\end{equation}

The resulting product is accumulated into the lower register over an effective length of \(n+m-1\). Because the final carry-out is omitted, the corresponding no-carry quantum-quantum adder has a \(T\)-depth of
\begin{equation}
n+m-2.
\end{equation}

For the second quantum-classical multiplication, the upper part lies within the effective \(m\)-bit range. Accumulating the remaining \(k-1\) partial products with no-carry adders gives
\begin{equation}
(m-1)(k-1).
\end{equation}

The second multiplication result is then accumulated using an \(n\)-bit no-carry adder with a \(T\)-depth of
\begin{equation}
n-1.
\end{equation}

The total folding \(T\)-depth is thus
\begin{equation}
\begin{aligned}
T_D^{\mathrm{fold}}
&=
n(k-1)
+
(n+m-2)
+
(m-1)(k-1)
+
(n-1) \\
&=
nk-n+n+m-2+mk-m-k+1+n-1 \\
&=
nk+mk-k+n-2 \\
&=
mk+(n-1)k+n-2.
\end{aligned}
\end{equation}

Using
\begin{equation}
k=\left\lfloor \frac{n}{2} \right\rfloor
\end{equation}
and
\begin{equation}
mk=\left\lfloor \frac{n^2}{4} \right\rfloor,
\end{equation}
we obtain
\begin{equation}
T_D^{\mathrm{fold}}
=
\left\lfloor \frac{n^2}{4} \right\rfloor
+
(n-1)\left\lfloor \frac{n}{2} \right\rfloor
+
n-2.
\end{equation}

For normalization, the additional sign qubit gives an effective register length of
\begin{equation}
p=n+1.
\end{equation}
The quantum-classical subtractor and controlled quantum-classical adder have \(T\)-depths of
\begin{equation}
p-1
\end{equation}
and
\begin{equation}
3p,
\end{equation}
respectively. The \(T\)-depth of one normalization block is therefore
\begin{equation}
\begin{aligned}
T_D^{\mathrm{norm1}}
&=
(p-1)+3p \\
&=
4p-1 \\
&=
4(n+1)-1 \\
&=
4n+3.
\end{aligned}
\end{equation}

Crandall reduction-2 applies this block twice, giving
\begin{equation}
T_D^{\mathrm{norm2}}
=
2T_D^{\mathrm{norm1}}
=
8n+6.
\end{equation}

The overall \(T\)-depth of Crandall reduction-1 is
\begin{equation}
\begin{aligned}
T_D^{\mathrm{CR1}}
&=
T_D^{\mathrm{fold}}
+
T_D^{\mathrm{norm1}} \\
&=
\left[
\left\lfloor \frac{n^2}{4} \right\rfloor
+
(n-1)\left\lfloor \frac{n}{2} \right\rfloor
+
n-2
\right]
+
(4n+3) \\
&=
\left\lfloor \frac{n^2}{4} \right\rfloor
+
(n-1)\left\lfloor \frac{n}{2} \right\rfloor
+
5n+1.
\end{aligned}
\end{equation}

Similarly, the overall \(T\)-depth of Crandall reduction-2 is
\begin{equation}
\begin{aligned}
T_D^{\mathrm{CR2}}
&=
T_D^{\mathrm{fold}}
+
T_D^{\mathrm{norm2}} \\
&=
\left[
\left\lfloor \frac{n^2}{4} \right\rfloor
+
(n-1)\left\lfloor \frac{n}{2} \right\rfloor
+
n-2
\right]
+
(8n+6) \\
&=
\left\lfloor \frac{n^2}{4} \right\rfloor
+
(n-1)\left\lfloor \frac{n}{2} \right\rfloor
+
9n+4.
\end{aligned}
\end{equation}

Each CCX-level block has a \(T\)-count of 4 under the depth-optimized temporary logical-AND implementation. Because the counted blocks form a sequential chain without parallel CCX-level execution, their number is equal to the derived \(T\)-depth. The corresponding \(T\)-counts are therefore
\begin{equation}
\begin{aligned}
T_{\mathrm{count}}^{\mathrm{CR1}}
&=
4T_D^{\mathrm{CR1}} \\
&=
4\left\lfloor \frac{n^2}{4} \right\rfloor
+
4(n-1)\left\lfloor \frac{n}{2} \right\rfloor
+
20n+4, \\
T_{\mathrm{count}}^{\mathrm{CR2}}
&=
4T_D^{\mathrm{CR2}} \\
&=
4\left\lfloor \frac{n^2}{4} \right\rfloor
+
4(n-1)\left\lfloor \frac{n}{2} \right\rfloor
+
36n+16.
\end{aligned}
\end{equation}

These expressions are worst-case logical resource estimates over the admissible range of \(c\). Constants with fewer active bits may require fewer partial-product accumulations and therefore lower resource costs.


\section{Decoder burden}
\label{subsec:decoder_burden}

The circuit execution time includes the time required to process the final backlog accumulated up to the last \(T\)-layer. However, evaluating the classical control overhead of an FTQC system also requires consideration of the decoder burden accumulated throughout circuit execution. We therefore use the cumulative backlog as a metric for decoder burden. It is defined as
\begin{equation}
B_{\mathrm{cum}}(n)
=
\sum_{k=1}^{D_T(n)}
\Delta_k.
\label{eq:cumulative_backlog}
\end{equation}

When \(f\neq1\), the cumulative backlog is
\begin{equation}
B_{\mathrm{cum}}(n)
=
\Delta_1(n)
\frac{f^{D_T(n)}-1}{f-1},
\label{eq:cumulative_backlog_f}
\end{equation}
whereas, for \(f=1\), it becomes
\begin{equation}
B_{\mathrm{cum}}(n)
=
D_T(n)\Delta_1(n).
\label{eq:cumulative_backlog_f1}
\end{equation}

Although this quantity is not added directly to the wall-clock execution time, it represents the total syndrome-data burden processed by the classical decoder during algorithm execution. The cumulative backlog can therefore be used to characterize decoder utilization, syndrome-memory requirements, classical interconnect traffic, and real-time control overhead.

For the neural-network decoder, \(f=21.8\), and the cumulative backlog therefore increases rapidly with the \(T\)-depth. By contrast, for the Sparse Blossom decoder, \(f=1\), and the cumulative backlog increases in proportion to \(D_T(n)\Delta_1(n)\). Sparse Blossom consequently exhibits more stable behavior than the neural-network decoder in terms of both circuit execution time and classical decoder burden.

Among the compared circuits, general Barrett reduction produces the largest cumulative backlog, whereas Crandall reduction-1 produces the smallest, consistent with the execution-time results. This indicates that the lower \(T\)-depth of Crandall reduction reduces not only quantum execution time but also decoder synchronization requirements and the cumulative syndrome-processing burden.

%
%

\end{document}